\documentclass[12pt]{article}

\usepackage{graphics,epsfig}

\def\simle{\mathrel{\rlap{\raise 0.511ex \hbox{$<$}}{\lower 0.511ex
 \hbox{$\sim$}}}}
\newcommand{\bea}{\begin{eqnarray}}
\newcommand{\eea}{\end{eqnarray}}
\newcommand{\be}{\begin{equation}}
\newcommand{\ee}{\end{equation}}
\newcommand{\nn}{\nonumber}
\newcommand{\gev}{{\rm~GeV}}
\newcommand{\mev}{{\rm~MeV}}
\newcommand{\fm}{{\rm~fm}}
\newcommand{\msb}{\overline{\rm{MS}}}
\newcommand{\mbar}{\overline{m}}

\def\ors{a}
\def\rmi{b}
\def\rmii{c}
\def\rmiii{d}
\def\infn{e}
\def\hum{f}
\def\infni{g}

\begin{document}

\begin{titlepage}
{
\normalsize
\hfill \parbox{125mm}{LPT-Orsay/10-77, RM3-TH/10-9, ROM2F/2010/16, HU-EP-10/59}}\\[10mm]
\begin{center}
  \begin{Large}
    \textbf{Average up/down, strange and charm quark masses \\
            with $N_f=2$ twisted mass lattice QCD\unboldmath} \\
  \end{Large}
\end{center}

\begin{figure}[h]
  \begin{center}
    \includegraphics[draft=false]{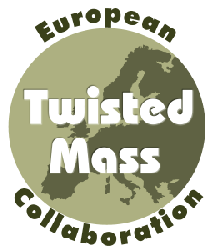}
  \end{center}
\end{figure}

\baselineskip 20pt plus 2pt minus 2pt
\begin{center}
  \textbf{
    B.~Blossier$^{(\ors)}$,
    P.~Dimopoulos$^{(\rmi)}$,
    R.~Frezzotti$^{(\rmii)}$,
    V.~Lubicz$^{(\rmiii,\infn)}$,\\
    M.~Petschlies$^{(\hum)}$,
    F.~Sanfilippo$^{(\rmi,\infni)}$,
    S.~Simula$^{(\infn)}$,
    C.~Tarantino$^{(\rmiii,\infn)}$
}\\
\end{center}

\begin{center}
  \begin{footnotesize}
    \noindent


$^{(\ors)}$ Laboratoire de Physique Th\'eorique (B\^at.~210), Universit\'e de
Paris XI,\\ Centre d'Orsay, 91405 Orsay-Cedex, France
\vspace{0.2cm}

$^{(\rmi)}$ Dipartimento di Fisica, Universit\`a di Roma ``Sapienza'', I-00185 Roma, Italy
\vspace{0.2cm}

$^{(\rmii)}$ Dip. di Fisica, Universit{\`a} di Roma Tor Vergata and\\ INFN, Sez.
di Roma Tor Vergata, Via della Ricerca Scientifica, I-00133 Roma, Italy
\vspace{0.2cm}

$^{(\rmiii)}$ Dip. di Fisica, Universit{\`a} Roma Tre, Via della Vasca Navale
84, I-00146 Roma, Italy
\vspace{0.2cm}

$^{(\infn)}$ INFN, Sez. di Roma
Tre, Via della Vasca Navale 84, I-00146 Roma, Italy
\vspace{0.2cm}




$^{(\hum)}$ Institut f\"ur Elementarteilchenphysik, Fachbereich Physik,
\\ Humboldt Universit\"at zu Berlin, D-12489, Berlin, Germany
\vspace{0.2cm}

$^{(\infni)}$ INFN, Sezione di Roma, I-00185 Roma, Italy


  \end{footnotesize}
\end{center}

\begin{abstract}
We present a high precision lattice calculation of the average up/down, strange
and charm quark masses performed with $N_f=2$ twisted mass Wilson fermions. The
analysis includes data at four values of the lattice spacing and pion masses as
low as $\simeq 270$ MeV, allowing for accurate continuum limit and chiral extrapolation.
The strange and charm masses are extracted by using several methods, based on
different observables: the kaon and the $\eta_s$ meson for the strange
quark and the $D$, $D_s$ and $\eta_c$ mesons for the charm. The quark mass
renormalization is carried out non-perturbatively using the RI-MOM method. The
results for the quark masses in the $\msb$ scheme read: $\mbar_{ud}(2\,\gev)=
3.6(2)\mev$, $\mbar_s(2\gev)=95(6)\mev$ and $\mbar_c(\mbar_c)=1.28(4)\gev$.
We also obtain the ratios $m_s/m_{ud}=27.3(9)$ and $m_c/m_s=12.0(3)$.

\end{abstract}
\end{titlepage}

\section{Introduction}
A precise knowledge of the values of the quark masses is of great importance for
testing the Standard Model (SM) of particle physics. From a phenomenological
point of view, several useful observables to constrain the SM or to search for
New Physics depend on quark masses, thus requiring accurate quark mass values in
order to allow for significant theory/experiment comparisons. From a more
theoretical side, explaining the quark mass hierarchy, which is not predicted by
the SM, is a deep issue and a great challenge. Lattice QCD calculations play a
primary role in the determination of quark masses. Recently, the progress
achieved thanks to several high statistics unquenched simulations is leading to
a significant reduction of the uncertainty on the quark mass values~\cite{Bazavov:2009bb}-\cite{Allison:2008xk}.

In this letter we present an accurate determination of the average up/down,
strange and charm quark masses performed by the European Twisted Mass
Collaboration (ETMC) with $N_f=2$ maximally twisted mass Wilson fermions. The
high precision of this analysis is mainly due to the extrapolation of the
lattice results to the continuum limit, based on data at four values of the
lattice spacing, to the well controlled chiral extrapolation, which uses
simulated pion masses down to $M_\pi \simeq 270 \mev$, and to the use of the
non-perturbative renormalization constants calculated
in~\cite{Constantinou:2010gr}. The only systematic uncertainty which is not
accounted for by our results is the one stemming from the missing strange and
charm quark vacuum polarization effects. Those are not accessible to us with
$N_f=2$ flavor simulations. However, a comparison of $N_f=2$ results for the up/down and
strange quark masses to already existing results from $N_f=2+1$ quark flavor
simulations~\cite{Scholz:2009yz} indicates that, for these observables, the error
due to the partial quenching of the strange quark is smaller at present than
other systematic uncertainties. 
The same conclusion is expected to be valid for the effects of the strange and charm partial quenching in the determination of the charm quark mass.

In this work, the calculation of the isospin averaged up/down quark mass, based on the study of
the pion mass and decay constant, closely follows the strategy
of~\cite{Baron:2009wt}. At variance with the latter, however, in the present
analysis we include data at four values of the lattice spacing, and use
the same lattice setup for all quark mass analyses.

For the strange quark mass, the main improvement
with respect to our previous work~\cite{Blossier:2007vv}, which used data at
a single lattice spacing only, is the continuum limit. Moreover, the
chiral extrapolation is performed by using either SU(2)- or SU(3)-Chiral
Perturbation Theory (ChPT). In order to extract the
strange quark mass we have used both the kaon mass and the mass of the (unphysical) $\eta_s$
meson composed of two degenerate valence strange quarks.
In both cases, the ultimate physical input is the kaon mass, together with the pion mass and decay constant.

For the charm quark mass, similarly to the strange quark, we have investigated
several experimental inputs to extract its value: the mass of the $D$, $D_s$ and
$\eta_c$ mesons.
In the charm quark sector discretization effects require some care and having data
at four lattice spacings helps in performing the continuum limit.
 
The results that we obtain for the quark masses are, in the $\msb$ scheme, 
\bea
&& \mbar_{ud}(2\gev) = 3.6(1)(2) \mev = 3.6(2)\mev\,,\nn\\
&&\mbar_s(2\gev) = 95(2)(6) \mev = 95(6)\mev\,,\nn\\
&&\mbar_c(\mbar_c) = 1.28(3)(3) \gev =1.28(4)\gev\,,
\label{eq:res1}
\eea
where the two separate errors are, respectively, statistical and systematic.
We also obtain for the ratios of quark masses the values
\bea
&&m_s/m_{ud}=27.3(5)(7)=27.3(9)\,,\nn\\
&&m_c/m_s=12.0(3)\,,
\label{eq:res2}
\eea
which are independent of both the renormalization scheme and scale.

\section{Simulation Details}
The calculation is based on the $N_f=2$ gauge field configurations generated by
the ETMC with the tree-level improved Symanzik gauge action~\cite{Weisz82}  and
the twisted mass quark action~\cite{FGSW01}  at maximal twist, discussed in detail
in~\cite{Baron:2009wt},\cite{Boucaud:2007uk}-\cite{Dimopoulos:2008sy}. We simulated $N_f=2$
dynamical quarks, taken to be degenerate in mass, whose masses are eventually
extrapolated to the physical isospin averaged mass of the up and down quarks.
As already mentioned, the strange and charm quarks are quenched in the present
calculation.

The use of the twisted mass fermions turns out to be beneficial, since the
pseudoscalar meson masses, which represent the basic ingredient of the
calculation, are automatically improved at ${\cal
O}(a)$~\cite{Frezzotti:2003ni}. As discussed
in~\cite{Blossier:2007vv,Frezzotti:2004wz,AbdelRehim:2006ve}, we implement
non-degenerate valence quarks in the twisted mass formulation by formally
introducing a twisted doublet for each non-degenerate quark flavor. In the
present analysis we thus include in the valence sector three twisted doublets,
($u,d$), ($s,s^\prime$) and ($c,c^\prime$), with masses $\mu_l$, $\mu_s$ and
$\mu_c$, respectively. Within each doublet, the two valence quarks are
regularized in the physical basis with Wilson parameters of opposite values,
$r=-r^\prime=1$.
Moreover, we only consider in the present study pseudoscalar mesons composed of valence quarks regularized with opposite $r$.
 This choice guarantees that the squared meson mass $m^2_{PS}$
differs from its continuum counterpart only by terms of ${\cal O}(a^2\,\mu_q)$
and ${\cal O}(a^4)$~\cite{Sharpe:2004ny,Frezzotti:2005gi}.

Details of the ensembles of gauge configurations used in the present analysis
and the values of the simulated valence quark masses are collected in
Tables~\ref{tab:simdet} and~\ref{tab:val}, respectively.
\begin{table}[t]
\begin{center}
\begin{tabular}{||c||c|c|c|c|c|c|c||} \hline 
 Ens.   &  $\beta$ &  $a\,[\fm]$ & $V/a^4$ & $a \mu_{sea}$ & $m_\pi\,[\mev]$ & 
$m_\pi \, L$ & $N_{cfg}$ \\
\hline\hline
$A_2$   & $3.8$ & $0.098$  &$24^3 \times 48$& $0.0080$   & $410$ & $5.0$ & $240$\\ 
$A_3$   &       &         &                & $0.0110$   & $480$ & $5.8$ & $240$\\ 
\hline
$B_1$   & $3.9$ & $0.085$ &$24^3 \times 48$& $0.0040$   & $315$ & $3.3$ & $480$\\ 
$B_2$   &       &         &                & $0.0064$   & $400$ & $4.1$ & $240$\\ 
$B_3$   &       &         &                & $0.0085$   & $450$ & $4.7$ & $240$\\ 
$B_4$   &       &         &                & $0.0100$   & $490$ & $5.0$ & $240$\\ 
\hline
$B_7$   & $3.9$ & $0.085$ &$32^3 \times 64$& $0.0030$   & $275$ & $3.7$ & $240$\\ 
$B_6$   &       &         &                & $0.0040$   & $315$ & $4.3$ & $240$\\ 
\hline
$C_1$   &$4.05$ & $0.067$ &$32^3 \times 64$& $0.0030$   & $300$ & $3.3$ & $240$\\ 
$C_2$   &       &         &                & $0.0060$   & $420$ & $4.5$ & $240$\\ 
$C_3$   &       &         &                & $0.0080$   & $485$ & $5.2$ & $240$\\ 
\hline
$D_1$   & $4.2$ & $0.054$ &$48^3 \times 96$& $0.0020$   & $270$ & $3.5$ & $80$\\ 
$D_2$   &       & $0.054$ &$32^3 \times 64$& $0.0065$   & $495$ & $4.3$ & $240$\\ 
\hline
\end{tabular}
\end{center}
\vspace{-0.4cm}
\caption{\sl Details of the ensembles of gauge configurations used in the
present study: value of the gauge coupling $\beta$; value of the
lattice spacing $a$; lattice size $V=L^3 \times T$ in lattice units; bare sea
quark mass in lattice units; approximate value of the pion mass; approximate
value of the product $m_\pi \, L$; number of independent configurations
$N_{cfg}$.}
\label{tab:simdet}
\end{table}
\begin{table}[h!]
\begin{center}
\begin{tabular}{||c||c|c|c||}
\hline
  &  $a \mu_l$ &   $a \mu_s$    & $a \mu_c$\\ \hline\hline
$A_2 - A_3$  & 0.0080, 0.0110 & 0.0165, 0.0200& 0.2143, 0.2406\\
             &                & 0.0250, 0.0300&   0.2701, 0.3032\\\hline
$B_1 - B_4$ & 0.0040, 0.0064,& 0.0150, 0.0180 & 0.2049, 0.2300\\
            &  0.0085, 0.0100 &   0.0220, 0.0270&  0.2582, 0.2898  \\ \hline
$B_7$ & 0.0030 & 0.0150, 0.0180  & 0.2049, 0.2300\\
       &        & 0.0220, 0.0270&  0.2582, 0.2898            \\\hline
$B_6$ & 0.0040 & 0.0150, 0.0180 & 0.2049, 0.2300\\
       &        & 0.0220, 0.0270&  0.2582, 0.2898            \\\hline
$C_1 - C_3$& 0.0030, 0.0060, & 0.0120,  0.0135& 0.1663, 0.1867 \\
           &  0.0080 &0.0150, 0.0180  &  0.2096, 0.2352\\\hline
$D_1$ & 0.0020 & 0.0130, 0.0150   & 0.1670, 0.1920  \\
      &        &0.0180 & 0.2170\\\hline
$D_2$ & 0.0065 & 0.0100, 0.0120 & 0.1700, 0.2200 \\
      &        &0.0150, 0.0190 &0.2700\\\hline

\hline
\end{tabular}
\end{center}
\vspace{-0.4cm}
\caption{\sl Values of simulated bare quark masses in lattice units for each
configuration ensemble in the light, strange and charm sectors.}
\label{tab:val}
\end{table}
In order to investigate the properties of the various light, strange and charmed mesons, we simulate the sea and valence up/down quark mass in the
range  $0.15\, m_s^{phys} \simle \mu_l \simle 0.5\, m_s^{phys}$, the valence
strange quark mass within $0.8\, m_s^{phys} \simle \mu_s \simle 1.5\,
m_s^{phys}$, and the valence charm quark mass within $0.9\, m_c^{phys} \simle
\mu_c \simle 2.0\, m_c^{phys}$, with $m_s^{phys}$ and $m_c^{phys}$ being the
physical strange and charm masses. Quark propagators with different valence masses
are obtained using the so called multiple mass solver method~\cite{Jegerlehner:1996pm,Jansen:2005kk}, which allows to invert the Dirac operator for several valence masses at a relatively low computational cost.

The statistical accuracy of the meson correlators is significantly improved by using the
so-called ``one-end" stochastic method, implemented in~\cite{McNeile:2006bz},
which includes all spatial sources at a single timeslice. Statistical errors on
the meson masses are evaluated using the jackknife procedure. With $16$
jackknife bins for each configuration ensemble we have verified that autocorrelations are well under control. Statistical errors on the fit
results which are based on data obtained from independent ensembles of gauge
configurations are evaluated using a bootstrap procedure, with $100$ bootstrap
samples, which properly takes into account cross-correlations.

The analysis is based on a study of the dependence of meson masses on
renormalized quark masses, with data at the four simulated values of the lattice
spacing simultaneously analyzed. For the quark mass renormalization constants 
$Z_{\mu}=Z_P^{-1}$ we use the results obtained in~\cite{Constantinou:2010gr},
which read
\be
Z_P\vert_\beta = \{0.411(12),\ 0.437(7),\ 0.477(6)\} \quad{\rm at}\quad
\beta=\{3.8,\ 3.9,\ 4.05\}\,,
\label{eq:ZP}
\ee
in the $\msb$ scheme at 2 GeV.
The errors given in eq.~(\ref{eq:ZP}) are those quoted in~\cite{Constantinou:2010gr} and do not account neither of discretization errors nor for the uncertainty associated to the perturbative conversion from the RI-MOM to the $\msb$ scheme.
The former are taken care by performing on the renormalized quark masses the extrapolation to the continuum limit.
The uncertainty associated to the conversion from the RI-MOM to the $\msb$ scheme is included in our final estimate of the systematic error on the quark masses and will be discussed in sect.~3.
For the renormalization constant at $\beta=4.2$, not calculated in~\cite{Constantinou:2010gr},
we use the preliminary result $Z_P^{\msb}(2\,\gev)|_{4.2}=0.501(20)$, where the conservative uncertainty is due
to the preliminarity of the result.

The uncertainty on $Z_P$ has been taken into account by
including in the definition of the $\chi^2$ to be minimized in the fits a
term
\be
\frac{\left( \widetilde Z_P^i(a)-Z_P^i(a) \right)^2}{\delta Z_P(a)^2}\,,
\ee
for each value of the lattice spacing and for each bootstrap sample, where
$Z_P^i(a)\pm \delta Z_P(a)$ is the input value for the renormalization constant
at the lattice spacing $a$ and for the bootstrap $i$, and $\widetilde Z_P^i(a)$
the corresponding fit parameter. This procedure corresponds to assuming for the
renormalization constant a Bayesian gaussian
prior~\cite{Baron:2009wt,Lepage:2001ym}.

The simultaneous analysis of data at different values of the lattice spacing
also requires the data conversion from lattice units to a common scale. For
the analysis in the pion sector, we have expressed all dimensionful quantities
in units of the Sommer parameter $r_0$~\cite{Sommer:1993ce}. We use for $r_0/a$
 in the chiral limit the values
\be
\left.\frac{r_0}{a}\right\vert_\beta = \{4.54(7),\ 5.35(4),\ 6.71(4),\
8.36(6)\} \quad{\rm at} \quad
\beta=\{3.8,\ 3.9,\ 4.05,\ 4.2\}\,,
\label{eq:r0a}
\ee
obtained from an extension of the analyses in~\cite{Baron:2009wt,Boucaud:2008xu}
with the inclusion of all four lattice spacings.
As in~\cite{Baron:2009wt}, the chiral extrapolation of $r_0/a$ is performed by
using three ans{\" a}tze for the sea quark mass dependence: linear only, quadratic only and quadratic+linear.
The size of mass-dependent discretization effects is verified by including in the fits $\mathcal{O}(a^2 m_l)$ and
$\mathcal{O}(a^2 m_l^2)$ terms, which turn out to be negligible. The uncertainties on the results given
in eq.~(\ref{eq:r0a}) include the systematic errors estimated as the spread among the values obtained from
the above-mentioned fits.
In the present analysis the uncertainty on the $r_0/a$
values is taken into account by adding a term to the $\chi^2$ of the fit in a similar way to $Z_P$, as explained above.

The analysis in the pion sector is also used to determine,
besides the value of the average up/down quark mass at the physical point, the
lattice spacing at each coupling $\beta$. The physical input used
for this determination is the pion decay constant $f_\pi$.
In the successive determination of the strange and charm quark masses,
the data are analyzed directly in physical units.

\section{Up/down quark mass}
The calculation of the up/down quark mass follows the strategy
of~\cite{Baron:2009wt}. The analysis is repeated here including simultaneously
all data available at the four values of the lattice spacing. 

We have studied the dependence of the pion mass and decay constant on the
renormalized quark mass.
For these quantities the predictions based on NLO ChPT and the Symanzik expansion up to $\mathcal{O}(a^2)$ can be written in  the form
\bea
m_\pi^2&=&(2\,B_0\,m_l) \cdot \left[1+\frac{2\,B_0\,m_l}{16 \pi^2\,f_0^2}\,\log \left(\frac{2\,B_0\,m_l}{16 \pi^2\,f_0^2}\right)+P_1\,m_l+a^2\cdot\left(P_2+P_3\,\log \left(\frac{2\,B_0\,m_l}{16 \pi^2\,f_0^2}\right)\right)  \right]\,,\nn\\
&&\nn\\
f_\pi&=&f_0\cdot \left[1-2\, \frac{2\,B_0\,m_l}{16 \pi^2 \,f_0^2}\,\log \left(\frac{2\,B_0\,m_l}{16 \pi^2 \,f_0^2}\right)
+ P_4\,m_l  + a^2\cdot\left(P_5+P_6\,\log \left(\frac{2\,B_0\,m_l}{16 \pi^2\,f_0^2}\right)\right)\,\right] \,,
\label{eq:mpifpi}
\eea
where $m_l$ is the renormalized light quark mass and $B_0$ and $f_0$ are the low energy constants (LECs) entering the LO chiral Lagrangian\footnote{The pseudoscalar decay constant in the chiral limit, $f_0$, is normalized such that $f_\pi=130.7 \mev$ at the physical point.}.

The coefficients of the discretization terms of $\mathcal{O}(a^2 m_l \log(m_l))$ receive a contribution from the $\mathcal{O}(a^2)$ splitting between the neutral and charged pion (squared) mass, $\Delta m^2_\pi = m^2_{\pi^0}-m^2_{\pi^\pm}$, which occurs with twisted mass fermions.
This contribution has been recently evaluated in~\cite{Bar:2010jk}.
Our main results are obtained through a fit of eqs.~(\ref{eq:mpifpi}), with the coefficients $P_3$ and $P_6$ obtained by expanding the results of~\cite{Bar:2010jk} up to $\mathcal{O}(a^2)$.
We have also verified that the results obtained in this way are indistinguishible from those obtained using directly the resummed formulae of~\cite{Bar:2010jk}.
From our fit, the splitting $\Delta m^2_\pi$ turns out to be determined with an uncertainty of approximately 60\%.
We obtain $\Delta m^2_\pi=- (33\pm19)\, a^2\, \Lambda_{QCD}^4$, which is consistent with a direct ETMC determination performed with two lattice spacings ($\Delta m^2_\pi=- 50\, a^2\, \Lambda_{QCD}^4$~\cite{Dimopoulos:2009qv}).
On the final result for the light quark mass the impact of this correction is at the level of the fitting error.

Lattice results for pion masses and decay constants have been corrected for finite size effects (FSE) evaluated using the resummed L\"uscher formulae.
The effect of the $\mathcal{O}(a^2)$ isospin breaking has been taken into account also in these corrections~\cite{Colangelo:2010cu}.
On our pion data, FSE vary between $0.2$\% and 2\%, depending on the simulated mass and volume.
The inclusion of the pion mass splitting in the FSE induces a variation of about 15-40\% in the finite size correction itself.
This effect is at the level of one third of the statistical error for our lightest pion mass at $\beta=3.9$ on the smaller volume, and even smaller in the other cases.

\begin{figure}[tb]
\includegraphics[width=0.4\textwidth,angle=270]{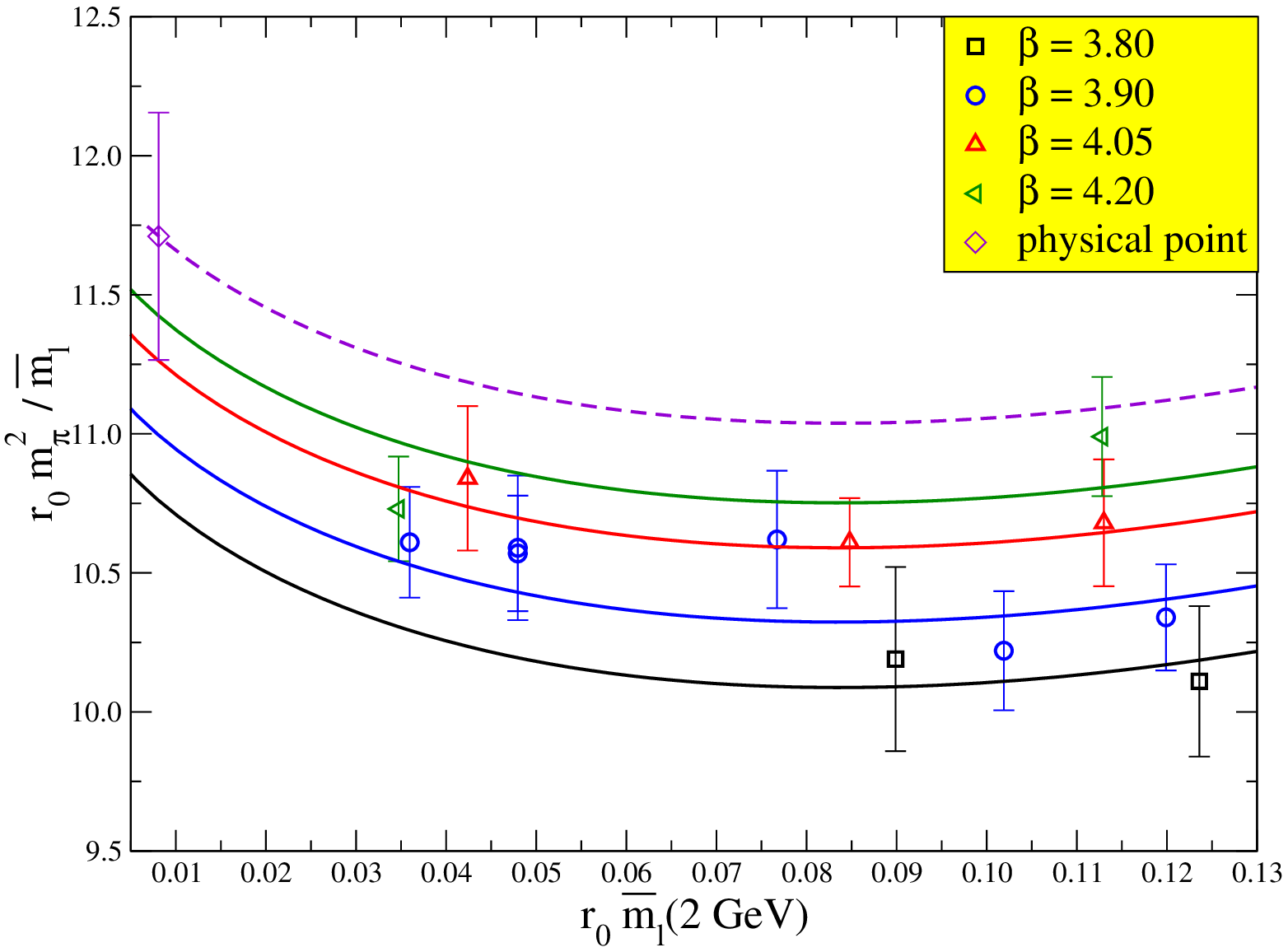}
\hspace{-0.8cm}
\includegraphics[width=0.42\textwidth,angle=270]{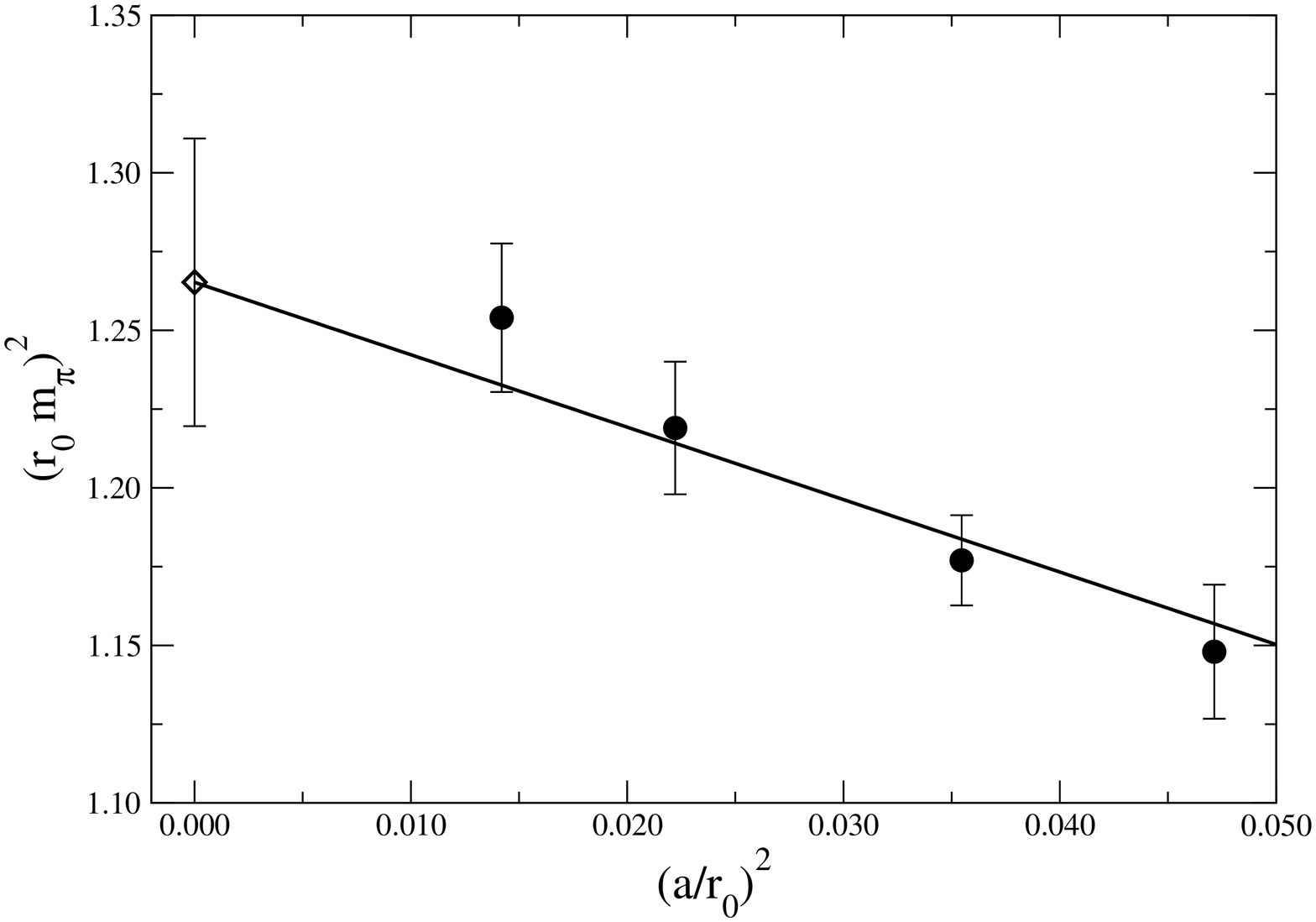}
\caption{\sl Left: Dependence of $r_0\,m_\pi^2/{\bar m}_l$ on the renormalized light quark mass at the four lattice spacings. Right: Dependence of $(r_0\,m_\pi)^2$ on the squared lattice spacing, for a fixed reference light quark mass ($\mbar_{ud}^{ref}=50\,\mev$). Empty diamonds represent continuum limit results.
\label{fig:mpi}}
\end{figure}

In fig.~\ref{fig:mpi} (left) we show the dependence of $r_0\,m_\pi^2/m_l$ on the renormalized light quark mass at the four $\beta$'s, and the curves corresponding to the best fit of the lattice data according to eq.~(\ref{eq:mpifpi}).

In order to illustrate the dependence of the pion mass on the lattice cutoff, we have interpolated the lattice data for $m_\pi^2$ at the four values of the lattice spacing to a common reference value of the light quark mass, $\mbar_{ud}^{ref}=50\,\mev$.
The resulting values of $(r_0\,m_\pi)^2$ obtained in this way are shown in fig.~\ref{fig:mpi} (right) as a function of $a^2$,
together with the corresponding continuum extrapolation.
We see that discretization errors on the pion mass square are below 10\% at $\beta=3.9$ and negligible within the fitting errors at $\beta=4.2$.

The value of the physical up/down quark mass is extracted from the ratio
$m_{\pi}^2/f_\pi^2$ using as an input the experimental value of the latter
ratio.\footnote{In order to account for the electromagnetic isospin
breaking effects which are not introduced in the lattice simulation, we have used in the present analysis as
``experimental" values of the pion and kaon mass the
combinations~\cite{Aubin:2004ck}
\be
(M_\pi^2)_{QCD} = M_{\pi^0}^2 \ ,
 \qquad (M_K^2)_{QCD} = \frac{1}{2} \left[
M_{K^0}^2 + M_{K^+}^2 - (1+\Delta_E) (M_{\pi^+}^2 - M_{\pi^0}^2) \right]\,,\nn
\ee
with $\Delta_E=1$.
The values of the experimental inputs for the pion and kaon masses are then $m_\pi^{exp}=135.0 \mev$, $m_K^{exp}=494.4 \mev$.
}
In order to estimate the systematic uncertainty due to discretization effects we have performed both a fit without the logarithmic discretization terms, i.e. with $P_3=P_6=0$ in eq.~(\ref{eq:mpifpi}) (the so called fit B of~\cite{Baron:2009wt}), and a fit without all $\mathcal{O}(a^2)$ corrections, i.e. with $P_2=P_3=P_5=P_6=0$ (the so called fit A of~\cite{Baron:2009wt}). Both these ans\"atze turn out be compatible with the lattice data.
We find that the result for the up/down quark mass decreases by approximately 2\% and increases of about 6\% in the two cases respectively, so that we estimate an overall uncertainty due to residual discretization effects of $\pm 4$\%.
We have also tried to add in the fit discretization terms of $\mathcal{O}(a^2 m_l^2)$ or $\mathcal{O}(a^4)$. In both cases these terms turn out to be hardly determined with our data, leading for $m_{ud}$ to results consistent with those obtained from the other fits, but with uncertainties larger by a factor three.

For estimating the systematic uncertainty due to the chiral extrapolation we have also considered a fit including a NNLO local contribution proportional to the light quark mass square. In this case we are not able to determine all the fitting parameters and we are thus forced to introduce, on the additional LECs, priors as in~\cite{Baron:2009wt}. In this way we find that the result for $m_{ud}$ increases by $6$\%.

The results of the fits described above are collected in Table~\ref{tab:ml} in the appendix.

As anticipated in the previous section, we also include in the final result a systematic uncertainty coming from the perturbative conversion of the quark mass renormalization constant from the RI-MOM to the $\msb$ scheme.
Using the results of the 3-loop calculation of~\cite{Chetyrkin:1999pq}, one can write the relation between the quark mass in the two schemes as
\be
\frac{\mbar(\mu)}{m^{\rm RI}(\mu)}= 1 - 0.424\, \alpha_s(\mu) - 0.827\, \alpha_s(\mu)^2 - 2.126\, \alpha_s(\mu)^3 +\mathcal{O}(\alpha_s(\mu)^4)\,.
\ee
The uncertainty due to the truncation of the perturbative series has been conservatively estimated by assuming the unknown $\mathcal{O}(\alpha_s^4)$ term to be as large as the $\mathcal{O}(\alpha_s^3)$ one.
Evaluating this term at the renormalization scale $\mu\simeq 3\,\gev$, which is the typical scale of the non-perturbative RI-MOM calculation in our simulation~\cite{Constantinou:2010gr}, and using $\alpha_s(3\,\gev, N_f=2)=0.202$, we then find that this uncertainty corresponds to approximately $\pm 2\,$\%. 

Adding in quadrature the three systematic errors discussed above we obtain $\mbar_{ud} = 3.55(14)(^{+28}_{-16}) \mev$ in the $\msb$ scheme at the renormalization scale of 2 GeV, where the two errors are statistical and systematic, respectively.
Finally we symmetrize the error, obtaining 
\be
\mbar_{ud}(2\,\gev) = 3.6(1)(2) \mev\,= 3.6(2) \mev\,.
\label{eq:rismud}
\ee
Note that, in the symmetrized result, the uncertainties due to discretization effects, chiral extrapolation and perturbative conversion give similar contributions to the final systematic error, at the level of 4\%, 3\% and 2\% respectively.

Using as an input the experimental value of the pion decay constant, the fits
also provide us with the values of lattice spacing at the four simulated
$\beta$'s, which are used in the rest of the analysis.
They read, at  $\beta=\{3.8, 3.9, 4.05, 4.2\}$ respectively,
\be
a|_\beta=\{0.098(3)(2), 0.085(2)(1), 0.067(2)(1),  0.054(1)(1)\}\,\fm\,,
\label{eq:afm}
\ee
where again the two errors are statistical and systematic.
The results in eqs.~(\ref{eq:rismud}) and~(\ref{eq:afm}) are in good agreement
with the previous ETMC determination obtained in~\cite{Baron:2009wt} from the analysis of data at $\beta=3.8, 3.9$ and $4.05$.

We observe that, in principle, the ratio of lattice spacings at two different
$\beta$ values could be determined from the fit of the pion meson mass and decay
constant, without using the additional information coming from the values of
$r_0/a$ of eq.~(\ref{eq:r0a}). With our data, however, the uncertainties on the values of the quark
mass renormalization constant, as well as the a priori unknown size of
discretization errors affecting the pion masses and decay constants, do not
allow to achieve a reliable determination of these ratios.

\section{Strange quark mass}
In this section, we first present the determination of the strange quark mass
based on the study of the kaon meson mass. The alternative determination based
on the study of the $\eta_s$ meson will be discussed later on.

Since the valence strange quark mass has not been previously tuned in our
simulation, the determination of the physical strange quark mass requires an
interpolation of the lattice data. As already mentioned, for all values of the
lattice couplings, the simulated values of the strange quark masses are
approximately in the range $0.8\, m_s^{phys} \simle \mu_s \simle 1.5\,
m_s^{phys}$.

In order to better discriminate the strange quark mass dependence of the kaon
masses on other dependencies, in particular discretization effects, we firstly
slightly interpolate the lattice data to three reference values of the strange quark
mass, which are chosen to be equal at the four lattice spacings: $\mbar_s^{ref} = \{ 80\,,\ 95\,,\ 110\} \mev$.
The interpolations to the reference
masses are performed by using quadratic spline fits. Then, at fixed reference
strange mass, we simultaneously study the kaon mass dependence on the up/down
quark mass and on discretization effects, thus performing the chiral
extrapolation and taking the continuum limit. In this step, we have considered
chiral fits based either on SU(2)-ChPT~\cite{Allton:2008pn,Roessl:1999iu}, where
the chiral symmetry is assumed for the up/down quark only, or partially
quenched SU(3)-ChPT~\cite{Sharpe:1997by}, where instead also the valence strange quark
is treated as light.
In order to extrapolate the kaon mass values to the continuum and to the physical
$m_{ud}$ limit, we use the results for the average up/down quark mass and for the
lattice spacings obtained in eqs.~(\ref{eq:rismud}) and (\ref{eq:afm}),
at each reference value of the strange quark mass.
Finally, we study the kaon mass dependence on the strange quark mass, and determine
the value of the physical strange quark mass using the experimental value of
$m_K$.

Let us describe the chiral fits in more detail. 
As discussed above, fits
are performed in two steps: 1) the strange quark mass is fixed to the
reference values and only the $m_l$ and $a^2$ dependence of the kaon mass is studied; 2) the
so obtained data in the continuum limit and at the physical $m_{ud}$ value are
studied as a function of the strange quark mass.
In these two steps, we have considered for the kaon
meson mass functional forms based on the predictions of either NLO SU(2)-ChPT~\cite{Allton:2008pn}, which predicts the absence at this order of chiral logs,
\bea
\label{eq:mK2SU2_1}
&&1)\, m_K^2(m_s, m_l, a)=Q_1(m_s)+Q_2(m_s)\,m_l+Q_3(m_s)\,a^2\,,\quad\forall\, m_s\,,\\
&&\nn\\
&&2)\, m_K^2(m_s, m_l^{phys}, a=0)\equiv Q_1(m_s)+Q_2(m_s)\,m_l^{phys}=Q_4+Q_5\, m_s\,,
\label{eq:mK2SU2_2}
\eea
or SU(3)-ChPT~\cite{Sharpe:1997by}
\bea
\label{eq:mK2SU3_1}
&&1)\, m_K^2(m_s, m_l, a)=B_0\cdot(m_s+m_l)\cdot(1+
Q_6(m_s)+Q_7(m_s)\, m_l+Q_8(m_s)\,a^2)\,,\,\forall\, m_s\,,\nn\\
&&\\
&&2)\, m_K^2(m_s, m_l^{phys}, a=0)\equiv B_0\cdot(m_s+m_l^{phys})\cdot(1+
Q_6(m_s)+Q_7(m_s)\, m_l^{phys})=\nn\\
&&\qquad\qquad =B_0\cdot(m_s+m_l^{phys})\cdot \left(1+\frac{2\,B_0\,m_s}{(4\pi f_0)^2}\,\log\frac{2\,B_0\,m_s}{(4\pi f_0)^2}+
Q_9\,m_s\right)\,,
\label{eq:mK2SU3_2}
\eea
where $B_0$ and $f_0$ are determined from the pion fit described in the previous
section.
Note that the dependence of the kaon mass on the strange quark mass is not determined by the chiral symmetry in the SU(2) theory.
We find that, with our choice of three reference strange masses around the physical value, a linear fit as given in eq.~(\ref{eq:mK2SU2_2}) is perfectly adequate to describe the data.

\begin{figure}[tb]
\includegraphics[width=0.4\textwidth,angle=270]{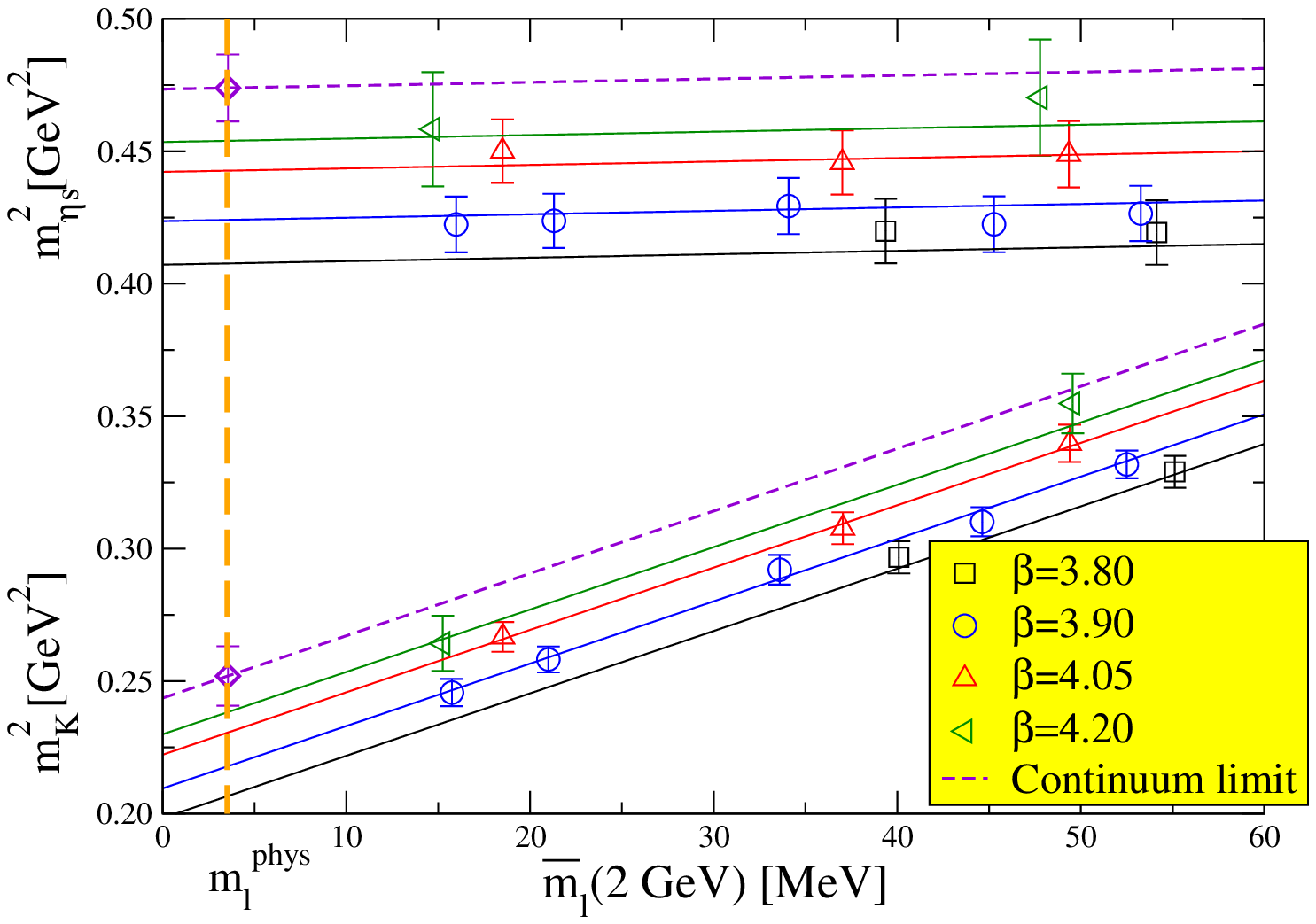}
\hspace{-0.8cm}
\includegraphics[width=0.38\textwidth,angle=270]{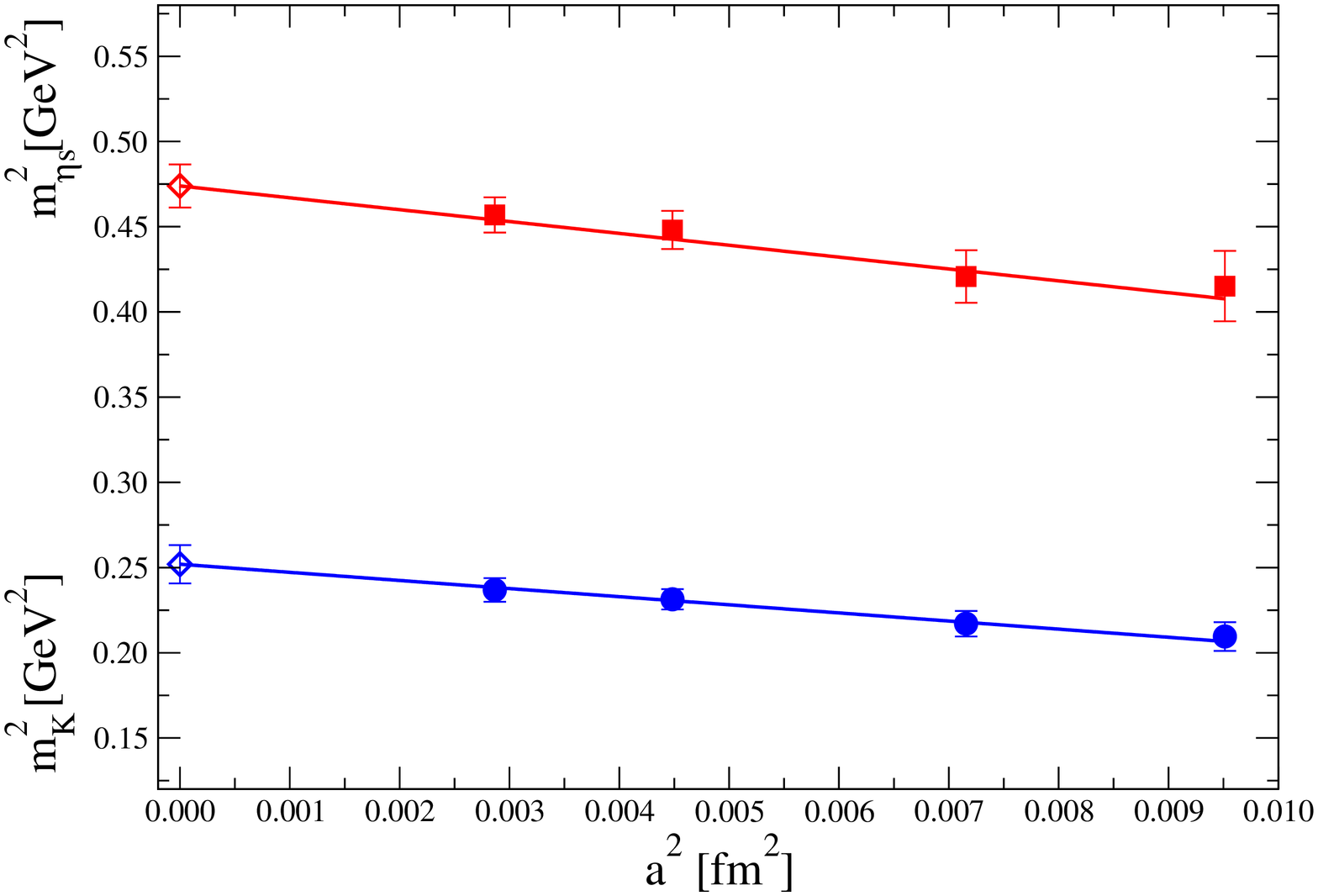}
\caption{\sl Left: Dependence of $m_K^2$ and $m_{\eta_s}^2$ on the renormalized light quark mass, for a
fixed reference strange quark mass ($\mbar_s^{ref}=95\,\mev$) and at the four lattice spacings. The orange vertical line corresponds to the physical up/down mass. Right: Dependence of $m_K^2$ and $m_{\eta_s}^2$ on the squared lattice spacing, for a fixed reference strange quark mass ($\mbar_s^{ref}=95\,\mev$) and at the physical up/down mass. Empty diamonds represent continuum limit results.
\label{fig:mKmss1}}
\end{figure}
\begin{figure}[tb]
\begin{center}
\includegraphics[width=0.45\textwidth,angle=270]{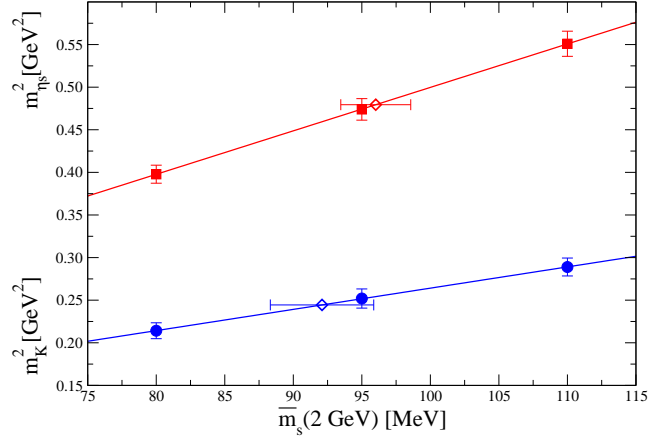} 
\caption{\sl Dependence of $m_K^2$ and $m_{\eta_s}^2$, in the continuum limit and at the physical up/down mass, on the strange quark mass. The strange mass results obtained from the SU(2)-ChPT analyses of kaon and $\eta_s$ mesons are also shown, with empty diamonds.
\label{fig:mKmss2}}
\end{center}
\end{figure}
For illustration we show in fig.~\ref{fig:mKmss1} the combined chiral/continuum fit based on SU(2)-ChPT, eq.~(\ref{eq:mK2SU2_1}), for a
fixed reference value of the strange quark mass, as a function of the light quark mass (left) and of the squared lattice spacing (right). In fig.~\ref{fig:mKmss2} the dependence on the strange quark mass is shown, in the case of the SU(2) analysis (see eq.~(\ref{eq:mK2SU2_2})). The dependencies are shown for the kaon squared mass as well as for the $\eta_s$ squared mass discussed hereafter.

As an alternative way to determine the strange quark mass we have studied the
dependence on $m_s$ of a meson made up of two strange valence
quarks~\cite{Davies:2009ih}. The advantage of this approach is that the mass of 
this unphysical meson, denoted as $\eta_s$, is only sensitive to the up/down quark mass through sea
quark effects, and it is thus expected to require only a very smooth chiral
extrapolation. This expectation will be confirmed by our analysis.
The price to pay is the need for an additional chiral fit required to determine the $\eta_s$ mass at the physical point.

In the real world, the $\eta_s$ meson is known to mix with the $(\bar uu +
\bar dd)$ component to produce the physical $\eta$ and $\eta'$ mesons.
This mixing proceeds through the contribution of disconnected diagrams,
which are known to be rather noisy on the lattice and therefore computationally expensive.
In order to avoid this computation we consider here the two strange quarks composing the meson as degenerate in mass but distinct in flavor. Though this $\eta_s$ meson does not exist in nature, its mass can be determined on the lattice~\cite{Davies:2009ih}.

In order to relate the mass of the $\eta_s$ meson to the physically observable $m_\pi$ and $m_K$, we have studied its dependence on the kaon and
pion masses for different values of the simulated light and strange quark
masses. This dependence turns out to be well described by both the
functional form\footnote{The functional forms in eqs.~(\ref{eq:etaKpiSU2})-(\ref{eq:etaKpiSU3}) are obtained from the ChPT formulae given in
eqs.~(\ref{eq:mssSU2_1})-(\ref{eq:mssSU3_2}) by replacing quark masses in terms of meson masses, and
keeping terms up to NLO.} based on either NLO SU(2)-ChPT,
\be
m_{\eta_s}^2=R_1+R_2\,(2\,m_K^2-m_\pi^2)+R_3\,m_\pi^2+R_4\,a^2\,,
\label{eq:etaKpiSU2}
\ee
or SU(3)-ChPT 
\bea
m_{\eta_s}^2&=&(2\,m_K^2-m_\pi^2)\cdot\left[1 + (\xi_s-\xi_l)\,\log(2\,\xi_s)+
(R_7+1)\,(\xi_s-\xi_l)+R_8\,a^2\right]\nn\\
&&- m_\pi^2\,\left[-\xi_l\,\log(2\,\xi_l)+\xi_s\,\log(2\,\xi_s)+R_7\,(\xi_s-\xi_l)\right]\,,
\label{eq:etaKpiSU3}
\eea
with $\xi_l=m_\pi^2/(4\,\pi\,f_0)^2$ and $\xi_s=(2\,m_K^2-m_\pi^2)/
(4\,\pi\,f_0)^2$.
We observe that, within the accuracy of our lattice data, the $\mathcal{O}(a^2)$ term in the $\eta_s$ mass is found to be independent of the strange quark mass. 

Once the physical values of the kaon and pion mass are inserted in eqs.~(\ref{eq:etaKpiSU2}) and~(\ref{eq:etaKpiSU3}), we find that the two fits yield very close results for the $\eta_s$ meson mass, namely
\be
m_{\eta_s} = 692(1) \mev  \quad {\rm from~SU(2)} \,, \qquad
m_{\eta_s} = 689(2) \mev  \quad {\rm from~SU(3)} \,,
\label{eq:mss}
\ee
to be compared with the LO SU(3) prediction $\left(m_{\eta_s}\right)_{LO}=\sqrt{2\,m_K^2 - m_\pi^2}=
686\mev$ and with the lattice determination of~\cite{Davies:2009ih} $m_{\eta_s}=686(4)\,\mev$.

Once the mass of the $\eta_s$ meson has been determined, the strange
quark mass can be extracted by following the very same procedure described for
the case of the kaon mass.
At first, lattice data at fixed reference strange mass are extrapolated to the continuum and to the physical up/down mass
(see fig.~\ref{fig:mKmss1}). After this extrapolation, the value of the physical strange quark mass is extracted by studying the dependence on the strange mass (see fig.~\ref{fig:mKmss2}).
We have considered the following fitting functions based on NLO-ChPT for the
dependence of the $\eta_s$ meson 1) on the (sea) up/down quark mass and on the
leading discretization effects, and 2) on the strange quark mass:
\bea
\label{eq:mssSU2_1}
&1)&\,m_{\eta_s}^2(m_s, m_l, a)=T_1(m_s)+T_2(m_s)\,m_l+T_3(m_s)\,a^2\,,\quad \forall\,m_s\,,\\
&&\nn\\
&2)&\,m_{\eta_s}^2(m_s, m_l^{phys}, a=0)\equiv T_1(m_s)+T_2(m_s)\,m_l^{phys}=T_4+T_5\,m_s\,,
\label{eq:mssSU2_2}
\eea
in SU(2), and 
\bea
\label{eq:mssSU3_1}
&1)&\, m_{\eta_s}^2(m_s, m_l, a)=2\,B_0\,m_s\cdot(1+T_6\,(m_s)+T_7(m_s)\,m_l+T_8(m_s)\,a^2)\,,\quad \forall\,m_s\,,\\
&&\nn\\
&2)&\, m_{\eta_s}^2(m_s, m_l^{phys}, a=0)\equiv 2\,B_0\,m_s\cdot(1+T_6\,(m_s)+T_7(m_s)\,m_l^{phys})=\nn\\
&&\qquad\qquad\qquad\qquad =2\,B_0\,m_s\cdot \left(2\,\frac{2\,B_0\,m_s}{(4\pi f_0)^2}\,\log(2\,\frac{2\,B_0\,m_s}{(4\pi f_0)^2})+T_9+T_{10}\,m_s \right)\,,
\label{eq:mssSU3_2}
\eea
in SU(3).
The LECs $T_2$ and $T_7$, describing the dependence on the light quark mass, are found to be independent of the strange mass, within the accuracy of our lattice data. They are then fitted with a single parameter for all reference strange quark masses.

The results for the strange quark mass obtained from both the kaon and the $\eta_s$ meson masses turn out to be well consistent, as it can be seen from Table~\ref{tab:ms} in the appendix where the several $m_s$ values obtained from different fits are collected.

In order to quote a final estimate for the strange quark mass we choose as a central value the weighted average of the results from the four determinations discussed above, namely from $K$ and $\eta_s$ and based on SU(2)- and SU(3)-ChPT.
In the $\msb$ scheme at 2 GeV this average reads $\mbar_s(2\,\gev)=95(2)\,\mev$, with the 2 MeV error representing the typical statistical and fitting uncertainty.
The difference between the determinations based on the $K$ and $\eta_s$ mesons is about 3\%.
The results obtained from either the SU(2) or the SU(3) fits are practically the same in the analysis based on the $\eta_s$ and differ by approximately 3\% in the kaon case.
In order to evaluate the uncertainty of the continuum extrapolation we have proceeded in two ways.
We have either added an $\mathcal{O}(a^4)$ mass independent term in eqs.~(\ref{eq:mK2SU2_1}),~(\ref{eq:mK2SU3_1}),~(\ref{eq:mssSU2_1}) and~(\ref{eq:mssSU3_1}), or we have excluded from these fits the data from the coarser lattice (with $a\simeq 0.098 \fm$).
We find that the $\mathcal{O}(a^4)$ term turns out to be hardly determined in the fit, leading to a factor three larger uncertainties.
The exclusion of $\beta=3.8$ data, instead, yields a variation of the results of approximately 2\% leaving the fitting error approximately unchanged.
We then assume $\pm2$\%  as uncertainty related to the continuum extrapolation.
The different fits considered for the determination of the up/down mass and of the lattice spacing affect the determination of the strange mass at the level of 3\%.
Finally, we include also in this case an uncertainty of 2\% related to the truncation of the perturbative expansion in the conversion from the RI-MOM to the $\msb$ scheme.
Combining all these uncertainties in quadrature, we quote as our final
estimate of the strange quark mass in the $\msb$ scheme
\be
\mbar_s(2\,\gev)=95(2)(6)\,\mev=95(6)\,\mev\,.
\label{eq:risms}
\ee

We observe that our result for the strange mass in eq.~(\ref{eq:risms}) is, though compatible,
smaller than the value obtained in~\cite{Blossier:2007vv} at a fixed value of
the lattice spacing ($a \simeq 0.085 \fm$). This is a consequence of discretization effects, which are at
the level of $15$\% in $m_K^2$ on the $a \simeq 0.085 \fm$ lattice, as shown in
fig.~\ref{fig:mKmss1} (right). 
A further comparison can be done with the ETMC estimate of the strange quark mass that appeared in the recent work on the bag parameter $B_K$~\cite{BK}.
Within that analysis, based on data at three $\beta$ values (3.8, 3.9 and 4.05), the strange quark mass is determined from the same lattice setup by performing an SU(2) chiral fit of the kaon meson mass. The result obtained in~\cite{BK} reads $\mbar_s(2\,\gev)=92(5)\,\mev$, to be compared to our result $\mbar_s(2\,\gev)=92.1(3.8)\,\mev$, obtained from the same fit (see Table~\ref{tab:ms}).  

Using our determinations of both the strange and light quark masses, we can obtain a prediction for the ratio $m_s/m_{ud}$, which is both a scheme and scale independent quantity. The several $m_s/m_{ud}$ values obtained from different fits are collected in Table~\ref{tab:mssumud} in the appendix.\footnote{The results for the ratio $m_s/m_{ud}$ collected in Table~\ref{tab:mssumud} are slightly different from the ratios of the $\mbar_s$ and $\mbar_{ud}$ results. This difference originates from the fact that in the ratio $m_s/m_{ud}$ the quark mass renormalization constant $Z_P^{-1}$ exactly cancels out, whereas in the determinations of $\mbar_s$ and $\mbar_{ud}$ the central values of  $Z_P$ are slightly modified by the fitting procedure.}
The result that we quote as our final estimate is
\be
m_s/m_{ud}=27.3(5)(7)=27.3(9)\,.
\ee

\section{Charm quark mass}
The determination of the charm quark mass follows, quite closely, the strategy
adopted in the determination of the strange quark mass discussed in the
previous section. In this case, we use as experimental input the masses of the 
$D$, $D_s$ and $\eta_c$ mesons. 

As for the strange quark case, the analysis requires an interpolation of the
lattice data, being the simulated charm masses roughly in the range $0.9\,
m_c^{phys} \simle \mu_c \simle 2.0\, m_c^{phys}$. In order to better
study the $a^2$ and $m_l$ dependence of charmed meson masses, we first use a
quadratic spline fit to interpolate the data at three reference 
values of the charm mass which are equal at the four $\beta$ values: $\mbar_c^{ref}(2\gev) = \{ 1.08\,,\ 1.16\,,\ 1.24\} \gev$.
We have verified that a different choice of the values of the reference masses leaves the charm quark results unchanged.
At fixed reference charm mass, we then study the dependence of the $D$, $D_s$ and
$\eta_c$ meson on the up/down mass (and on the strange mass in the case of the
$D_s$ meson) and on discretization terms, thus getting the results for the meson
masses in the continuum limit, at the physical values of the light (and strange) quark masses,
and at the reference charm mass. Finally, the value of the physical charm quark
mass is extracted by fitting these data as a function of the charm quark mass
and using as an input the experimental value of the corresponding charmed meson
$m_D^{exp}= 1.870\,\gev$, $m_{D_s}^{exp}=1.969\,\gev$, $m_{\eta_c}^{exp}=2.981\,\gev$.\footnote{The experimental value of the meson masses should be corrected to take into account the absence of electromagnetic effects and, in the case of the $\eta_c$, of disconnected diagrams in the lattice calculation. For the $\eta_c$ meson, these corrections are estimated to be of the order of 5 MeV~\cite{Davies:2009ih}, thus affecting the extracted charm quark mass to approximately $0.2$\%. Similar corrections are expected for the $D$ and $D_s$ mesons. Given our uncertainties, we can safely neglect these corrections in the analysis.}

In order to fit the meson masses we have considered the following (phenomenological) polynomial fits, which turn out
to describe well the dependence on the light and strange quark masses and on the lattice cutoff of the $D$, $D_s$ and
$\eta_c$ meson masses, at fixed (reference) charm mass $m_c$,
\be
m_{H}(m_c,m_s,m_l,a)=C_1^{H}(m_c)+C_2^{H}(m_c)\,m_l+C_3^{H}(m_c)\,m_s+C_4^{H}(m_c)\,a^2\,,\quad \forall m_c\,,
\label{eq:DDsetac_ref}
\ee
with $H=D_, D_s, \eta_c$.
From the fits, we find that the coefficients $C_2^{H}$ and $C_3^{H}$
turn out to be independent of the charm mass within the statistical errors.
The latter coefficient $C_3^{H}$, of course, enters the fit only in the $D_s$ case.

For the charm mass dependence, a constant plus either a linear or a $1/m_c$ term have been considered for describing data of the $D$, $D_s$ and $\eta_c$ mesons, namely
\bea
m_{H}(m_c,m_s^{phys},m_l^{phys},a=0)&\equiv& C_1^{H}(m_c)+C_2^{H}(m_c)\,m_l^{phys}+C_3^{H}(m_c)\,m_s^{phys}=\nn\\
&=&C_5^{H}+\frac{C_6^{H}}{m_c}+C_7^{H}\,m_c\,.
\label{eq:mhq_mc}
\eea
Since we have data at three reference charm masses (close to the physical charm), we can keep only one of the coefficients $C_6^{H}, C_7^{H}$ different from zero. We find that both choices describe very well the lattice data and affect only in a marginal way the interpolation to the physical charm mass.

\begin{figure}[tb]
\includegraphics[width=0.38\textwidth,angle=270]{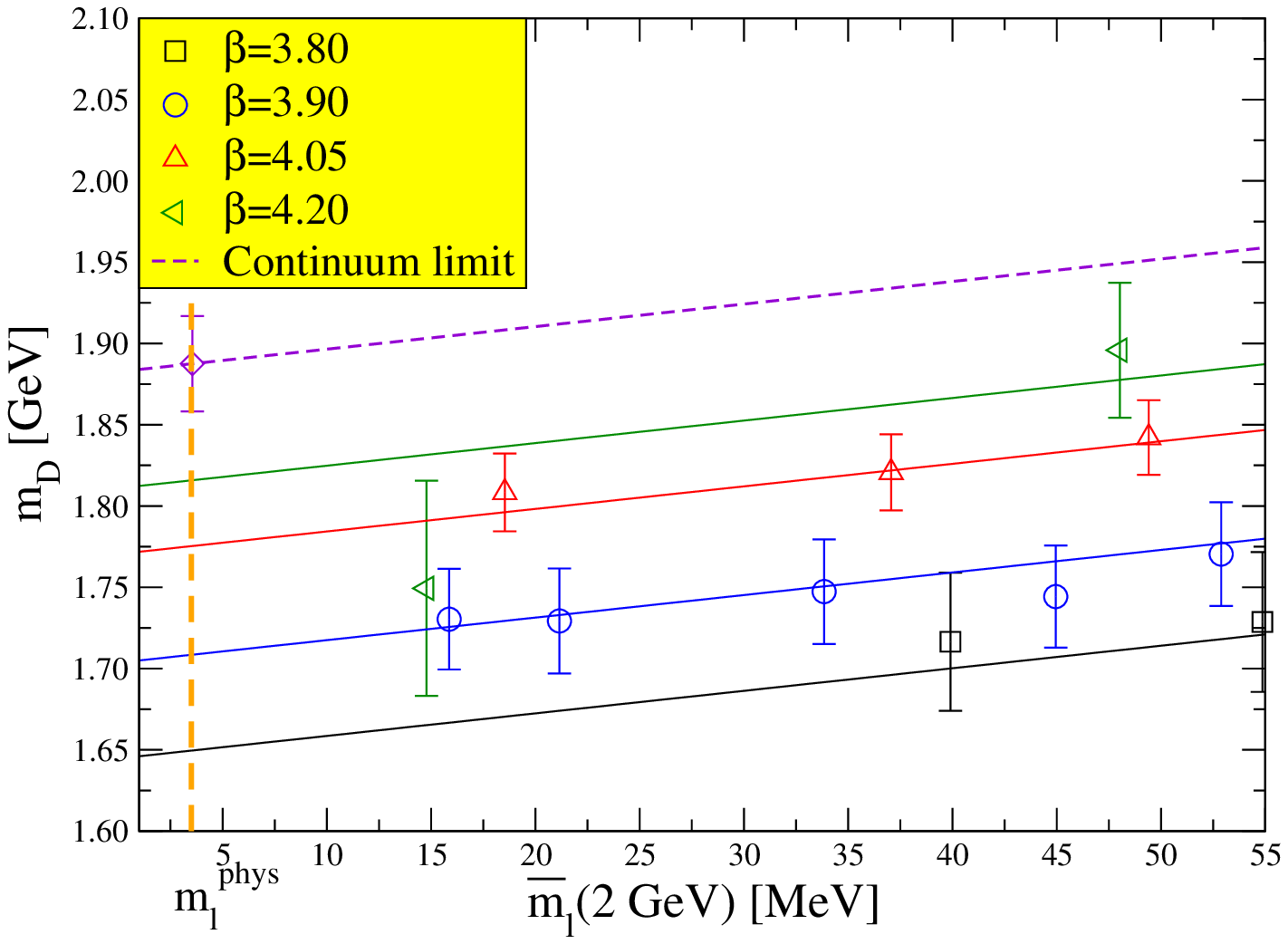}
\hspace{-0.8cm}
\includegraphics[width=0.38\textwidth,angle=270]{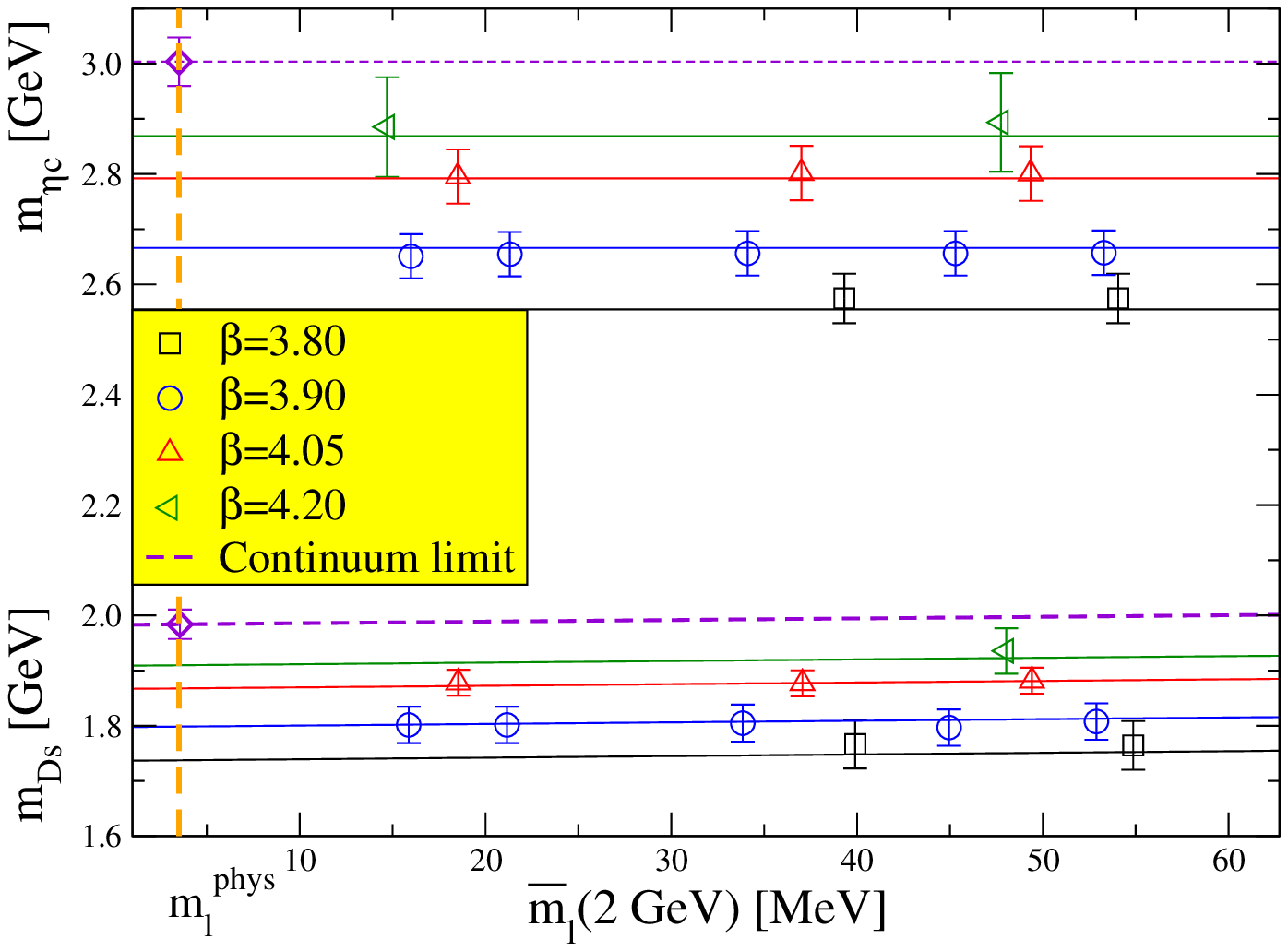} 
\caption{\sl Left: Dependence of $m_D$ (left) and  $m_{D_s}$ and $m_{\eta_c}$ (right) on the light quark mass, at
fixed reference charm quark mass ($\mbar_c^{ref}=1.16\,\gev$) and for the four simulated
lattice spacings. For the $D_s$ meson the strange quark mass is fixed to the reference value $\mbar_s^{ref}=95 \,\mev$. 
\label{fig:mDmDsmetac_vs_ml}}
\end{figure}

In fig.~\ref{fig:mDmDsmetac_vs_ml} we show the dependence of the $D$, $D_s$ and $\eta_c$ masses on the light quark mass at a fixed reference charm mass, for the four $\beta$'s. For the $D_s$ and $\eta_c$ mesons, which contain the light
quark in the sea only, this dependence turns out to be practically invisible.

\begin{figure}[tb]
\includegraphics[width=0.38\textwidth,angle=270]{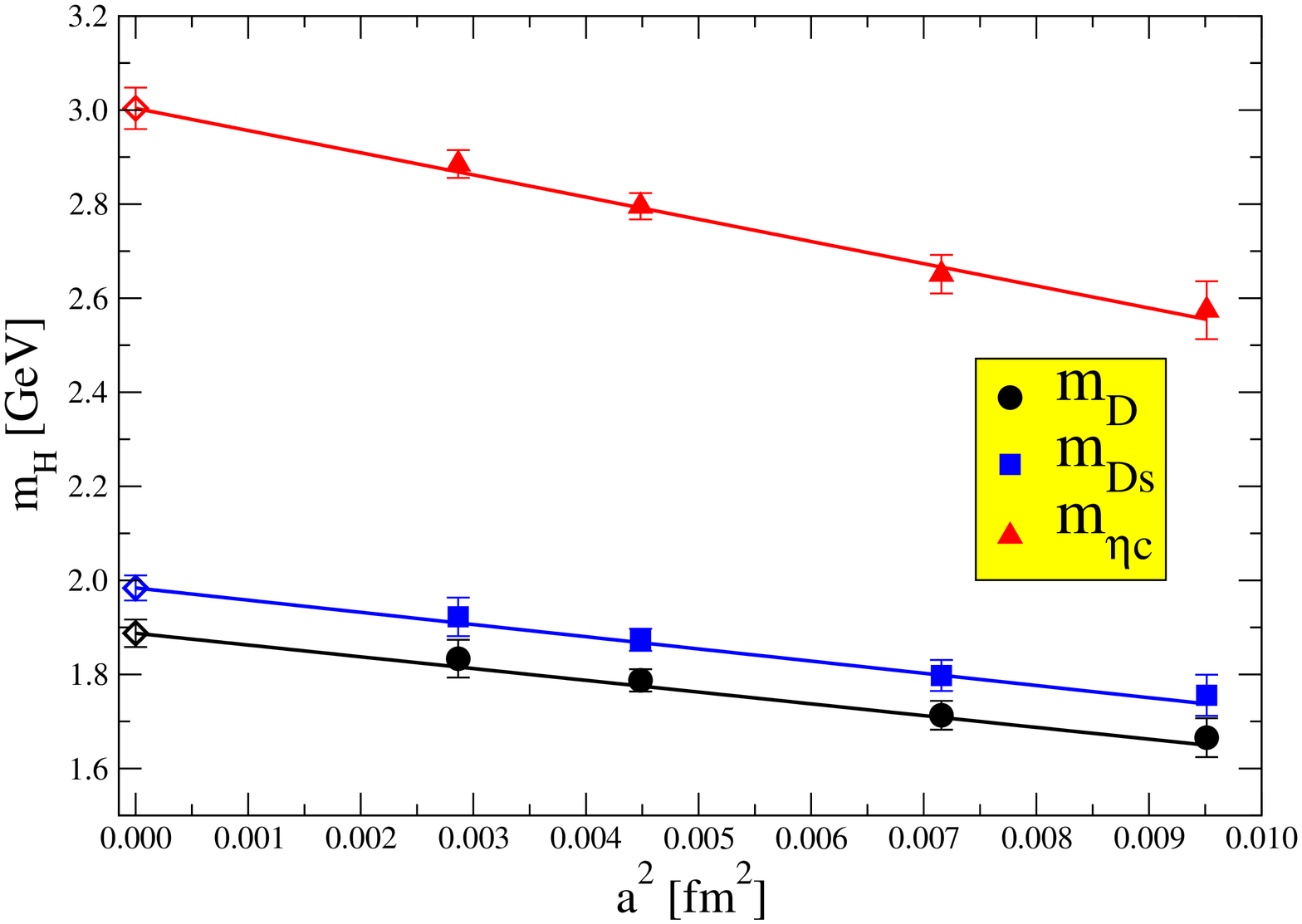}
\hspace{-0.8cm}
\includegraphics[width=0.38\textwidth,angle=270]{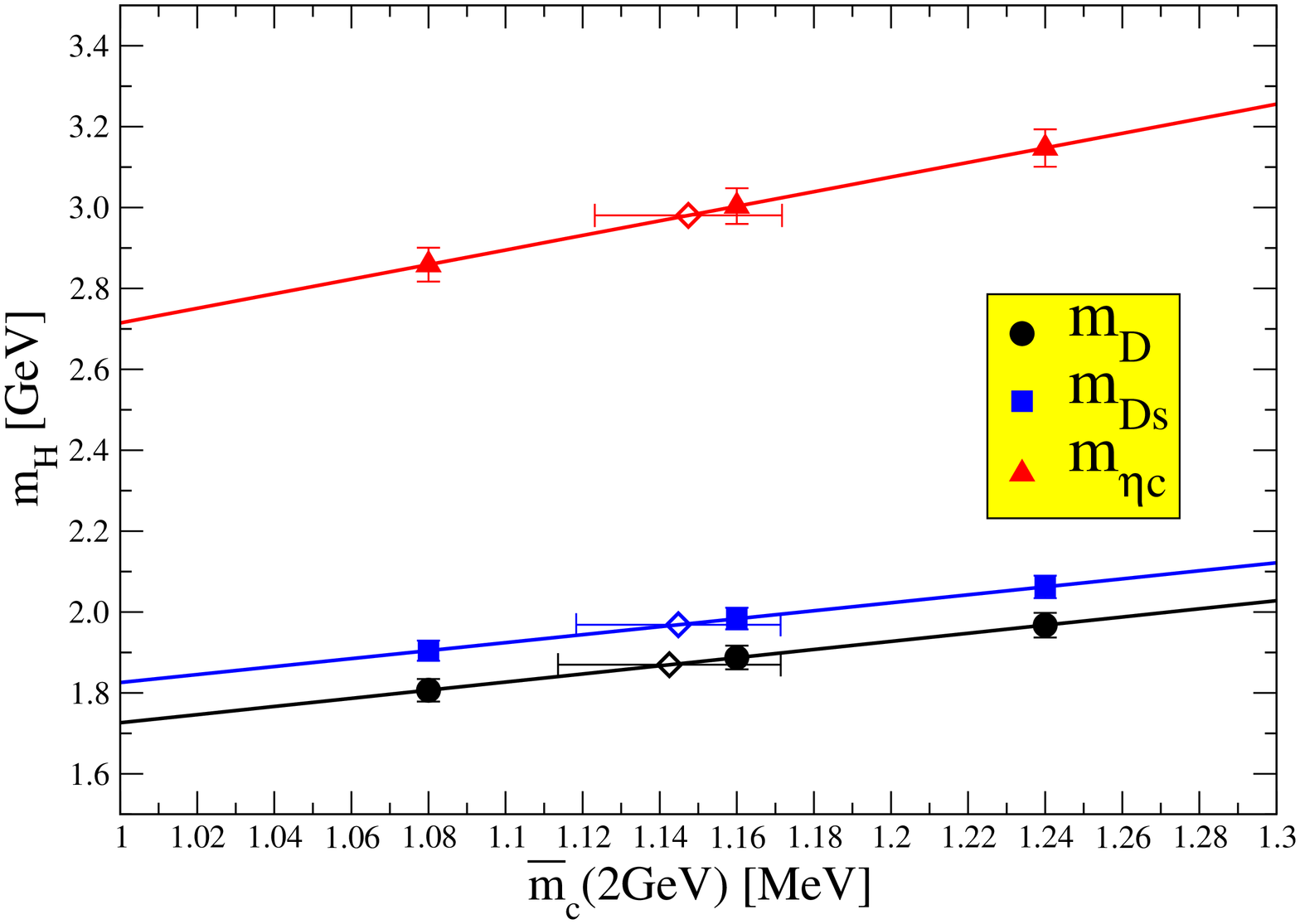} 
\caption{\sl Left: Dependence of $m_D$, $m_{D_s}$ and $m_{\eta_c}$, at
fixed reference charm quark mass ($\mbar_c^{ref}=1.16\,\gev$) and at physical up/down and strange quark mass, on the squared lattice spacing. Empty diamonds represent continuum limit results. Right: Dependence of $m_D$, $m_{D_s}$ and $m_{\eta_c}$, in the continuum limit and at physical up/down and strange quarks, on the charm quark mass. The charm mass results from the three determinations are also shown, with empty diamonds.
\label{fig:mDmDsmetac_vs_a2mc}}
\end{figure}

In fig.~\ref{fig:mDmDsmetac_vs_a2mc} (left) the meson masses at  physical light and strange quark masses are shown as a function of $a^2$, for a reference value of the charm quark mass.
As can be seen from this plot, discretization effects on the $\eta_c$ meson mass vary from approximately 4\% on the finest lattice up to 14\% on the coarsest one. These effects are larger than those affecting the $D$ and $D_s$ meson masses by approximately 30\%.
Fig.~\ref{fig:mDmDsmetac_vs_a2mc} (left) also shows that the dependence of the three charmed meson masses on $a^2$ is very well described by a linear behaviour, and attempts to vary the continuum extrapolation with respect to the simple linear fit produce only small effects. The latter are included in the estimate of the systematic uncertainty, as discussed below.

Finally, fig.~\ref{fig:mDmDsmetac_vs_a2mc} (right) shows the dependence of the $D$, $D_s$ and $\eta_c$ masses on the charm mass (in the continuum limit and at physical light and strange mass) and the interpolation to the physical charm.

Using the experimental values of the considered meson masses we get for the charm quark mass the results collected in Table~\ref{tab:mc}.

In order to estimate the uncertainty due to the continuum extrapolation we have proceeded in two ways.
We have either added in the fitting form of eq.~(\ref{eq:DDsetac_ref}) an $\mathcal{O}(a^4)$ dependence, which turns out to be hardly determined thus leading to uncertainties larger by a factor three, or we have excluded the data from the coarser lattice (with $a\simeq 0.098 \fm$). This latter analysis yields a variation of the results of approximately $1.5$\%.
The two dependencies of the meson masses on the charm quark mass, considered in eq.~(\ref{eq:mhq_mc}), yield results that differ by only few MeV.
The systematic uncertainty then comes from the sum in quadrature of the approximately 1\% spread among the three determinations from the $D$, $D_s$ and $\eta_c$ mesons, the $1.5$\% uncertainty due to discretization effects and the 2\% uncertainty coming from the the perturbative conversion of the renormalization constants from the RI-MOM to the $\msb$ scheme.

We quote as our final result for the charm quark mass in the $\msb$ scheme
\bea
\label{eq:rismc}
\mbar_c(2 \gev)&=&1.14(3)(3) \gev=1.14(4) \gev\,\\
\rightarrow ~~ \mbar_c(\mbar_c)&=&1.28(4) \gev\,,\nn
\eea
where the evolution to the more convential scale given by
$\mbar_c$ itself has been performed at N$^3$LO~\cite{Chetyrkin:1999pq} with $N_f=2$, consistently with our non-perturbative evaluation of the renormalization constant. Had we evolved with $N_f=4$, which is the number of active flavours above $\mu=m_c$, the result for $\mbar_c(\mbar_c)$ would have increased by less than one standard deviation.

Our result is compatible with the preliminary estimate of the charm quark mass, $\mbar_c(2 \gev)=1.23(6)$, obtained by ETMC~\cite{Blossier:2009hg} using data at three lattice spacings and preliminary values for the renormalization constants.
It is also in good agreement with the HPQCD result $\mbar_c(\mbar_c)=1.268(9)\,\gev$~\cite{Allison:2008xk}, with a larger uncertainty in our determination.
Finally, our result is in good agreement with the recent sum rules determination $\mbar_c(\mbar_c)=1.279(13)\,\gev$ of~\cite{Chetyrkin:2009fv}.

We also provide a prediction for the scheme and scale independent ratio $m_c/m_s$. The several $m_c/m_s$ values obtained from different fits are collected in Table~\ref{tab:mcsums} in the appendix.
The result that we quote as our final estimate is
\be
m_c/m_{s}=12.0(3)\,,
\ee
in good agreement with the other recent lattice determination $m_c/m_s=11.85(16)$~\cite{Davies:2009ih}.

\section{Conclusions}
We have presented results for the average up/down, strange
and charm quark masses, obtained with $N_f=2$ twisted mass Wilson fermions. The
analysis includes data at four values of the lattice spacing and pion masses as
low as $\simeq 270$ MeV, allowing a well controlled continuum limit and chiral extrapolation.
Within the strange sector the
chiral extrapolation is performed by using either SU(2)- or SU(3)-ChPT.
The strange and charm masses are extracted by using several methods, based on
different meson mass inputs: the kaon and the $\eta_s$ meson for the strange
quark and the $D$, $D_s$ and $\eta_c$ mesons for the charm. The quark mass
renormalization is carried out non-perturbatively using the RI-MOM method.

The results for the quark masses in the $\msb$ scheme read: $\mbar_{ud}(2\,\gev)=
3.6(2)\mev$, $\mbar_s(2\gev)=95(6)\mev$  and $\mbar_c(\mbar_c)=1.28(4)\gev$.
The quoted errors include the uncertainty in the perturbative conversion of the renormalization constants from the RI-MOM to the $\msb$ scheme, which is conservatively estimated to be at the level of 2\%.
We emphasize that this uncertainty is not related to the lattice calculation itself, but comes from continuum perturbation theory.
If the RI-MOM scheme was chosen as a reference scheme and, say, 3 GeV as a reference scale, which is the typical scale of the non-perturbative RI-MOM calculation in our lattice simulation~\cite{Constantinou:2010gr}, this uncertainty would not be present at all.
For reference we provide our results for the quark masses also in this scheme: $m^{\rm{RI}}_{ud}(3\gev) = 3.9(1)(2) \mev$, $m^{\rm{RI}}_s(3\gev) = 102(2)(6) \mev$ and $m^{\rm{RI}}_c(3\gev) = 1.22(3)(2) \gev$.

We have also evaluated the quark mass ratios $m_s/m_{ud}=27.3(9)$ and $m_c/m_s=12.0(3)$, which are independent on both the renormalization scale and scheme.

The only systematic uncertainty which is not
accounted for by our results is the one stemming from the missing strange and
charm quark vacuum polarization effects. A comparison, for instance, of our $N_f=2$ result for the
strange quark mass, to already existing results from $N_f=2+1$ quark flavor
simulations~\cite{Scholz:2009yz} indicates that the error
due to the partial quenching of the strange quark is smaller at present than
other systematic uncertainties.
In this respect we mention that simulations with $N_f=2+1+1$ dynamical flavors are already being performed by ETMC and preliminary results for several flavor physics observables have been recently presented~\cite{Baron:2010bv,Baron:2010th}.

\vspace{0.3cm}
We thank all the ETMC members for fruitful discussions and the apeNEXT computer 
centres in Rome for their invaluable technical help.
Some computation time has been used for that project on the
BlueGene system at IDRIS.
We are also grateful to Gilberto Colangelo for having provided us with a routine for the calculation of the FSE~\cite{Colangelo:2010cu}, and to Chris Sachrajda for drawing our attention on  the uncertainty in the perturbative conversion of the renormalization constants.

\section{Appendix}
We collect in this appendix the results obtained for the up/down, strange and charm quark masses from the different fits considered in the present analysis.

As discussed in sect.~3, in the light quark sector we have performed the following fits:
\begin{itemize}
\item[-] L1: this is our best fit which is based on NLO ChPT with the inclusion of $\mathcal{O}(a^2)$ discretization effects. This fit corresponds to eq.~(\ref{eq:mpifpi}) with all parameters different from zero.
\item[-] L2: same as L1 but without discretization terms, i.e. $P_2=P_3=P_5=P_6=0$ in eq.~(\ref{eq:mpifpi}).
\item[-] L3: same as L1 with the inclusion of a NNLO correction proportional to the square of the light quark mass.
\end{itemize}
The results for the up/down quark mass obtained from these fits are collected in Table~\ref{tab:ml}.
For illustration, we also show in the table the value of $\mbar_{ud}$ obtained from a fit (denoted as L4 here and B in~\cite{Baron:2009wt}) without logarithmic discretization terms, i.e. with $P_3=P_6=0$ in eq.~(\ref{eq:mpifpi}), and without isospin breaking corrections in the FSE.
\begin{table}[ht!]
\begin{center}
\begin{tabular}{||c||c|c|c|c||}
\hline
& L1    & L2 & L3 & L4\\ \hline\hline
$\mbar_l\, [\mev]$ & $3.55(14)$  & $3.75(7)$ & $3.78(17)$ & $3.47(11)$ \\
\hline\hline
\end{tabular}
\end{center}
\vspace{-0.4cm}
\caption{\sl Results for the up/down quark mass in the $\msb$ scheme at 2 GeV, as obtained from fits L1, L2, L3 and L4.}
\label{tab:ml}
\end{table}

For the strange quark mass  we collect the results of the different fits in Table~\ref{tab:ms}, where we use the short notation: K-SU(2), K-SU(3), $\eta_s$-SU(2), and $\eta_s$-SU(3) for distinguishing the determinations from the kaon and $\eta_s$ masses, and based on SU(2)- and SU(3)-ChPT.

\begin{table}[ht!]
\begin{center}
\begin{tabular}{||c||c|c|c|c||}
\hline
$\mbar_s\, [\mev]$& K-SU(2) & K-SU(3) & $\eta_s$-SU(2) & $\eta_s$-SU(3) \\ \hline\hline
L1 & $92.1(3.8)$ & $94.7(2.2)$ & $96.0(2.6)$ & $95.5(2.1)$\\\hline
L2 & $91.6(3.9)$ & $94.6(2.3)$ & $95.4(2.6)$ & $95.3(1.9)$\\\hline
L3 & $95.4(3.8)$ & $94.7(2.1)$ & $99.4(2.9)$ & $97.7(2.2)$\\
\hline\hline
\end{tabular}
\end{center}
\vspace{-0.4cm}
\caption{\sl Results for the strange quark mass in the $\msb$ scheme at 2 GeV, as obtained from the different fits within the light and strange quark sectors.}
\label{tab:ms}
\end{table}
The results for the ratio $m_s/m_{ud}$ are given in Table~\ref{tab:mssumud}.
\begin{table}[ht!]
\begin{center}
\begin{tabular}{||c||c|c|c|c||}
\hline
$m_s/m_{ud}$& K-SU(2) & K-SU(3) & $\eta_s$-SU(2) & $\eta_s$-SU(3) \\ \hline\hline
L1 & $26.9(5)$ & $27.2(5)$ & $27.6(4)$ & $27.3(7)$\\\hline
L2 & $27.1(5)$ & $26.9(3)$ & $27.5(3)$ & $26.8(3)$\\\hline
L3 & $25.7(5)$ & $26.0(6)$ & $26.5(6)$ & $26.0(7)$\\
\hline\hline
\end{tabular}
\end{center}
\vspace{-0.4cm}
\caption{\sl Results for the ratio $m_s/m_{ud}$, as obtained from the different fits within the light and strange quark sectors.}
\label{tab:mssumud}
\end{table}

Finally, for the charm quark mass and the ratio $m_c/m_s$ the results are collected in Tables~\ref{tab:mc} and~\ref{tab:mcsums} , where $D$, $D_s$ or $\eta_c$ indicate the meson whose mass is used as input.
These analyses are practically insensitive to the choice of the fit in the pion sector and only the results obtained from the fit L1 are shown in the tables. Similarly, for the ratio $m_c/m_s$ the values shown in Table~\ref{tab:mcsums} correspond to the analysis of the $D$ meson only, since the analyses of the $D_s$ or $\eta_c$ mesons yield practically identical results.
\begin{table}[ht!]
\begin{center}
\begin{tabular}{||c||c|c|c||}
\hline
& $D$ & $D_s$ & $\eta_c$ \\ \hline\hline
$\mbar_c\, [\gev]$ & $1.14(3)$ & $1.14(3)$ & $1.15(2)$ \\
\hline\hline
\end{tabular}
\end{center}
\vspace{-0.4cm}
\caption{\sl Results for the charm quark mass in the $\msb$ scheme at 2 GeV, as obtained from the different fits within the charm sector.}
\label{tab:mc}
\end{table}
\begin{table}[ht!]
\begin{center}
\begin{tabular}{||c||c|c|c|c||}
\hline
& K-SU(2) & K-SU(3) & $\eta_s$-SU(2) & $\eta_s$-SU(3) \\ \hline\hline
$m_c/m_s$ & $12.4(4)$ & $12.1(2)$ & $11.9(2)$ & $12.0(3)$\\\hline\hline
\end{tabular}
\end{center}
\vspace{-0.4cm}
\caption{\sl Results for the ratio $m_c/m_s$, as obtained from the different fits within the strange quark sector, and from the analysis of the $D$ meson mass in the charm sector.
}
\label{tab:mcsums}
\end{table}

\end{document}